\documentclass[twocolumn,groupedaddressm,showkeys,byrevtex]{revtex4}
\usepackage{graphicx}
\usepackage{bm} 

\begin{document}

\newcommand{\Mfunction}[1]{#1}
\newcommand{\imag}{\imath}
\newcommand{\Mvariable}[1]{#1}


\title{Modeling Copper Cables and Connectors\footnote{\copyright 2002, All Rights Reserved}}

\author{Eric C. Hannah}
\email[]{eric.hannah@intel.com}
\affiliation{Intel Corporation \\
RNB6-37, 2200 Mission College Blvd., Santa Clara, CA 95052-8119}

\date{\today}

\begin{abstract}
High-speed data busses constructed from copper components are both difficult to model and important to accurately simulate. Using analytic signal theory and an extension of the skin effect for non-uniform conductivity, experimental data taken from IEEE 1394-1995 cables and connectors is used to predict the transfer function of a 4.5 meter cable with two connectors. With these techniques any length of cable between two connectors can be simulated with high accuracy, including all the cable and connector losses and the infinitely summed standing waves that occur between the connectors.
\end{abstract}


\maketitle

\preprint{\copyright 2002, All Rights Reserved}

\section{Introduction}       
Modern high-speed busses send data streams across long runs of copper, often at data rates in excess of 1 gigabit/second. Connectors at the ends of copper cables create discontinuities that phase shift the electrical signals and cause reflections. Mathematically characterizing these copper systems with high accuracy is challenging and often performed with awkward lumped-circuit models. In this paper I demonstrate methods that provide a new level of accuracy and convenience in modeling such systems. To demonstrate the accuracy of this method the IEEE 1394-1995 connector and cable system is analyzed for jitter and losses.

\section{Copper interconnect}
\begin{figure}
\includegraphics[scale=0.8]{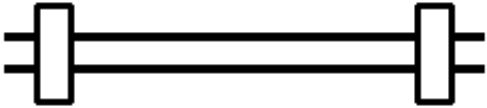}
\caption{\label{fig:CopperInter}Copper interconnect}
\end{figure}

Figure \ref{fig:CopperInter} shows a simple copper interconnect consisting of two connectors and a long run of two parallel copper wires.

We restrict the discussion to the differential mode of excitation, excluding common mode effects -- and also neglect cross-talk and EMI issues. The connectors have complex effects upon the phase and amplitude of each Fourier component of the electrical waves incident upon them. They cannot be treated as simple lumped elements in high-speed systems. Next, the wires have distributed effects upon the electrical waves and should not be treated as a transmission line with a simple skin depth effect -- real wires have non-uniform conductivity due to stresses and alloying effects in their manufacturing. Finally, the entire system acts as a lossy resonator with standing wave modes.

Accurate modeling and simulation of this simple a system is both difficult and necessary. High-speed cable systems may have frequency-dependent losses specified to the 1 dB level and jitter budgets defined to 100 ps -- for 5 meters of cable. Errors in the predicted waveforms are catastrophic in systems that may ship in the 100 million units per year volume.

\section{Analytic signals}
The analytic signal\cite{gabor} is a complex-number representation of a signal. The basic notion is that we start with a real-valued signal which has no explicit phase information, make this into the corresponding analytic signal, and thereby determine the phase factors of the complex signal. (See Appendix \ref{App:analytic}). Analytic signals are essential for proper modeling of copper interconnects.	

\subsection{Amplitude-Phase-Frequency ambiguity}
Following \cite{vakman} we write a signal in the form
\begin{eqnarray}
u(t) & = & a(t) \cos \phi (t) \nonumber \\
& = & a(t) \cos (\omega(t)  t + \Phi (t) ), \nonumber \\
& & u(t) \in \mathcal{R}
\end{eqnarray}
$a(t)$, $\Phi(t)$, and $\omega(t) \equiv d\phi/dt$ are the amplitude, phase, and frequency of the signal. However, for a real-valued signal $u(t)$, this is an equation relating two unknowns to a single known. It may be argued that given the smooth envelope of maximums and the zero-crossings we know what the amplitude and frequency are. To a certain extent this viewpoint is effective for narrowband signals. It is totally inapplicable for wideband signals.

It is advantageous to work with complex signals
\begin{equation}
w(t) = u(t) + \imag v(t) = a(t) e^{\imag \phi (t)}
\end{equation}
constructed by adding an imaginary part $v(t)$ to the real signal $u(t)$. For complex signals the amplitude, phase and fequency are well defined
\begin{equation}
a^2(t) = u^2 + v^2 = |w|^2
\end{equation}
\begin{equation}
\phi(t) = \mbox{arctan}[\frac{v}{u}]
\end{equation}
\begin{equation}
\omega (t) = \mbox{Im}[\frac{w^\prime}{w}]
\end{equation}

If $v(t)$ is the Hilbert transform of $u(t)$ then $w(t)$ is the analytic signal. The Hilbert transform is
\begin{equation}
H[u(t)] \equiv \frac{1}{\pi}\int _{-\infty} ^\infty \frac{u(s)}{t-s} ds
\end{equation}
Vakman{\cite{vakman} has demonstrated that only the Hilbert-transform imaginary part of the complex signal satisfies the following conditions: 1) amplitude continuity and differentiability under small perturbations, 2) phase-value independence of signal scaling, and, 3) harmonic correspondence, i.e., sine waves of constant frequency and amplitude retain their values.

The analytic signal provides a cogent and global definition of the amplitude, phase and frequency of a signal. The analytic signal is implicitly implemented in radio devices such as filters and modulators. Finally, the analytic signal provides accurate estimates of the amplitude, phase and frequency of wide-band signals even in the presence of substantial noise.

\subsection{Fourier transform for analytic signals}
To efficiently convert time-dependent real signals to analytic signals we use the Fourier transform.

\subsubsection{Continuous-time analytic signal}
Following Marple\cite{marple} let $x(t)$ be a real-valued, finite-energy signal defined over the temporal interval $-\infty < t < \infty$ with the corresponding continuous-time Fourier transform (CTFT)
\begin{equation}
X(f) = \int _{-\infty} ^\infty x(t) e^{(-\imag 2 \pi f t)} dt
\end{equation}
defined over the frequency interval $-\infty < f < \infty$. Because $x(t)$ is real, the CFTF is complex-conjugate symmetric, $X(-f) = X^*(f)$.

The continuous-time analytic signal $z(t)$ corresponding to $x(t)$ is most simply defined in the frequency domain as
\begin{equation}
Z(f) \equiv  \left\{ \matrix{2X(f),&  f > 0 \cr X(0),&  f = 0 \cr0,&  f < 0} \right.
\end{equation}
which is inverse transformed to obtain $z(t)$. Note that the value of $Z(f)$ at $f = 0$ is defined to reflect the mathematical behavior of the Fourier transform at a discontinuity, which yields the average of the values on either side of the discontinuity. It can be shown that the imaginary part of $z(t)$ is the Hilbert transform of the real signal $x(t)$.

\subsubsection{Discrete-Time "Analytic" signal properties}
Again following Marple\cite{marple} we consider the properties appropriate for an analytic-like discrete-time signal $z[n]$ corresponding to a real-valued discrete-time signal $x[n]$ of finite duration $T$.

The spectrum of the signal $x[n]$ is obtained from the discrete-time Fourier transform (DTFT)
\begin{eqnarray}
X(f) & = & T \sum _{n=0} ^{N - 1} x[n] e^{(-\imag 2 \pi f n T)}, \nonumber \\
&& f = 0, 1, \ldots, N-1
\end{eqnarray}
The DTFT is evaluated with a fast Fourier transform. Next form the $N$-point one-sided discrete-time "analytic" signal transform
\begin{equation}
Z(m) \equiv  \left \{ \matrix{X(0),&  m = 0 \cr 2X(m),&  1 \leq m \leq \frac{N}{2} - 1 \cr X[\frac{N}{2}],&  m = \frac{N}{2} \cr 0,&  \frac{N}{2} +1 \leq m \leq N -1} \right.
\end{equation}
Compute using an $N$-point inverse DTFT
\begin{equation}
z[n] \equiv \frac{1}{N T} \sum _{m=0} ^{N - 1} Z[m] e^{(+\imag 2 \pi m n / N)}
\end{equation}
to yield the complex discrete-time "analytic" signal for the same sample rate as the original signal $x[n]$.

One justification for setting the special value at $m = \frac{N}{2}$ is that 0 Hz and the Nyquist frequency terms are shared boundaries between negative and positive frequency halves of the periodic spectrum, and division into respective one-sided positive and negative spectra requires that these terms be split between the two spectral halves.

\section{Experimental methods}
We set up in the laboratory a standard IEEE 1394-1995 connector and cable assembly which was interfaced to a Tektronics Digital Sampling Scope with a Time Domain Reflectometry module (TDR). TDR data from production cables and connectors was analyzed for impedance discontinuities and then used to model data-dependent jitter. 

\subsection{Time domain reflectometry}
TDR directly measures the reflected voltage from a test fixture as a function of delay time. A step function with a very short risetime is launched into the connected system. After the launch trigger the input of the system is monitored for voltage levels. Voltage levels are affected by spatial variations in the impedance of the connected system, as changes in impedance scatter the outgoing waveform back towards the source point.

\subsection{Digital sampling oscilloscope operation}
The Tektronix 11801B oscilloscope contains a two-channel 20GHz bandwidth sampling head. The oscilloscope functions in equivalent-time sampling. A sampling interval is defined as the horizontal time divided by the record length (number of defined samples). As an example, for a sample interval of 200ps/div the oscilloscope advances by one sample interval (200ps) after each trigger event. Hence, for a record length of 512, it would take 512 trigger events for the oscilloscope to advance to the end of the screen. 

TDR operation adds a step recovery diode circuit that generates a step pulse with a 40 ps rise time into the Device Under Test at the 0 time strobe. All the reflected signals are sampled by the DSO system to show the time profile of the DUT's reflections.

\section{Experimental work}
The connector was a surface-mount style and the cable was a D7 style. Only one connector and cable assembly was tested though our results correspond closely to data given to us confidentially by other companies.

The Figure \ref{fig:TDRconnCable} shows the differential TDR signal for a production 1394-1995 connector and cable.
\begin{figure}
\includegraphics*[scale=0.4]{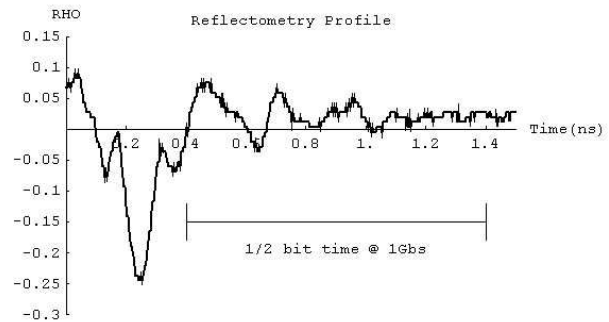}
\caption{\label{fig:TDRconnCable}TDR of connector and cable}
\end{figure}
Figure \ref{fig:TDRconn} shows the TDR data for a connector with the cable removed.
\begin{figure}
\includegraphics*[scale=0.4]{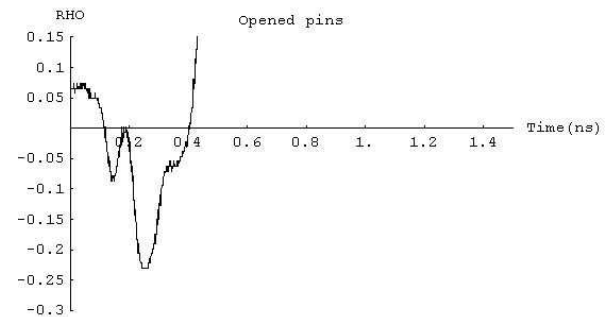}
\caption{\label{fig:TDRconn}TDR of connector with cable disconnected}
\end{figure}
We see there is a substantial dip in the connector's impedance.

It is useful to convert TDR data into an equivalent impedance profile. For differential TDR the conversion formula is:
\begin{equation}
Z = 100 \, \Omega \, (\frac{1+\rho}{1 - \rho})
\end{equation}
Note that the cable impedance in Figure \ref{fig:Zprofile} is above the 100 Ohms level (about 105 Ohms). The ideal 1394 cable differential impedance is 110 Ohms.
\begin{figure}
\includegraphics*[scale=0.4]{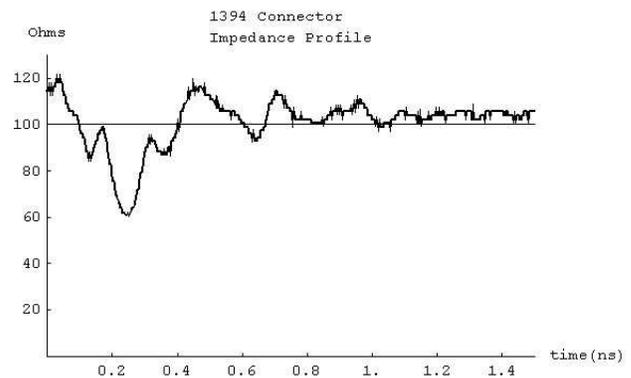}
\caption{\label{fig:Zprofile}Equivalent impedance profile}
\end{figure}

\subsection{Complex impedance of the connector}
Following the numerical algorithm given above we now create the analytic signal representation of Figure \ref{fig:Zprofile}'s impedance profile (which has no phase information). This algorithm will generate the imaginary component of the impedance.
\begin{figure}
\includegraphics*[scale=0.4]{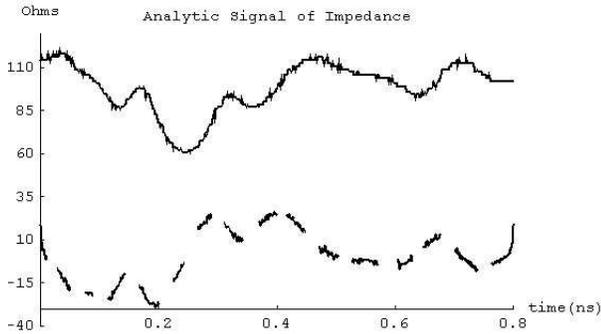}
\caption{\label{fig:CmpZprofile}Equivalent impedance profile}
\end{figure}
Figure \ref{fig:CmpZprofile} is the resulting analytic signal. The top line is the real part of the analytic signal and the bottom, broken line, is the imaginary part. Taken as a complex function this diagram is the complex impedance profile of the connector. A lumped-circuit approximation for this would require a series of capacitors and inductors to match each minimum and maximum in the trace -- a large number of components. Additionally, such a series of components would have secondary reflections that would need to be added into the fitting procedure so that the total, consistent, electrical system matched the TDR data.

\subsection{S-parameters}
We now want to derive the frequency-dependent S-parameters for transmission through and reflectance off the connector.

The definition of the transmission coefficient, $S_{12}$, is:
\begin{equation}
S_{12}(\omega) = T(\omega) = \frac{2 Z_{\mbox{load}}(\omega)}{Z_{\mbox{load}}(\omega)+Z_0(\omega)}
\end{equation}
Similarly the definition of the reflection coefficient $S_{11}$ is:
\begin{equation}
S_{11}(\omega) = R(\omega) = \frac{Z_{\mbox{load}}(\omega)- Z_0(\omega)}{Z_{\mbox{load}}(\omega)+Z_0(\omega)}
\end{equation}

The most straightforward way to derive these coefficients for the connector is to create a synthetic pulse with an exactly known Fourier transform (see Figure \ref{fig:synPulse}), use the time profile of the TDR derived connector complex impedance to calculate reflectance (see Figure \ref{fig:refPulse} -- magnitude only) and transmission (see Figure \ref{fig:transPulse} -- magnitude only) synthetic pulse patterns, then Fourier transform the time profile of these pulses to get the analytic spectra. One subtlety is to use the variation of the measured impedance from the nominal 110$\Omega$ value of the transmission line, not the absolute impedance. It is only the variations from the uniform value that scatters waves into the backward direction. For accuracy we use the actual transmission line impedance value measured (by suitable averaging away from the connector). A further subtlety is that we have implicitly summed up all the internal reflections inside the connector by using the TDR data.
\begin{figure}
\includegraphics*[scale=0.4]{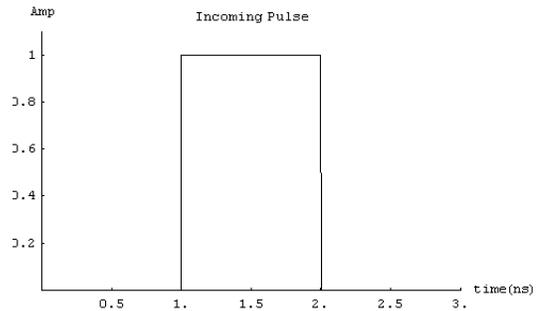}
\caption{\label{fig:synPulse}Synthetic pulse}
\end{figure}
\begin{figure}
\includegraphics*[scale=0.4]{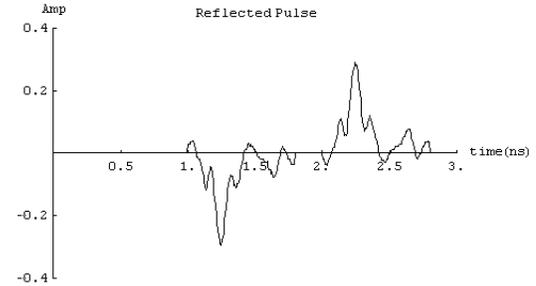}
\caption{\label{fig:refPulse}Reflected pulse}
\end{figure}
\begin{figure}
\includegraphics*[scale=0.4]{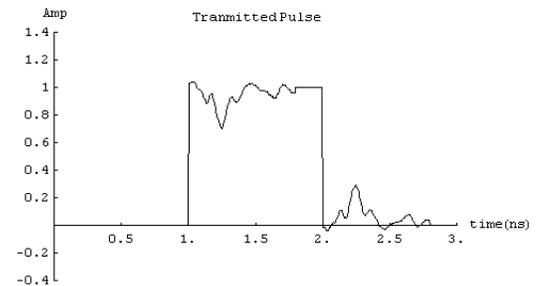}
\caption{\label{fig:transPulse}Transmitted pulse}
\end{figure}

Finally, we divide the analytic spectrum of the reflected and transmitted pulses by the analytic spectrum of the original pulse to get the complex ratio at each frequency, which is the transfer ratio as a function of frequency. This procedure simulates using a Network Analyer. Figure \ref{fig:S12} is the S matrix function for $S_{12}$, Figure \ref{fig:S11} is $S_{11}$.
\begin{figure}
\includegraphics*[scale=0.4]{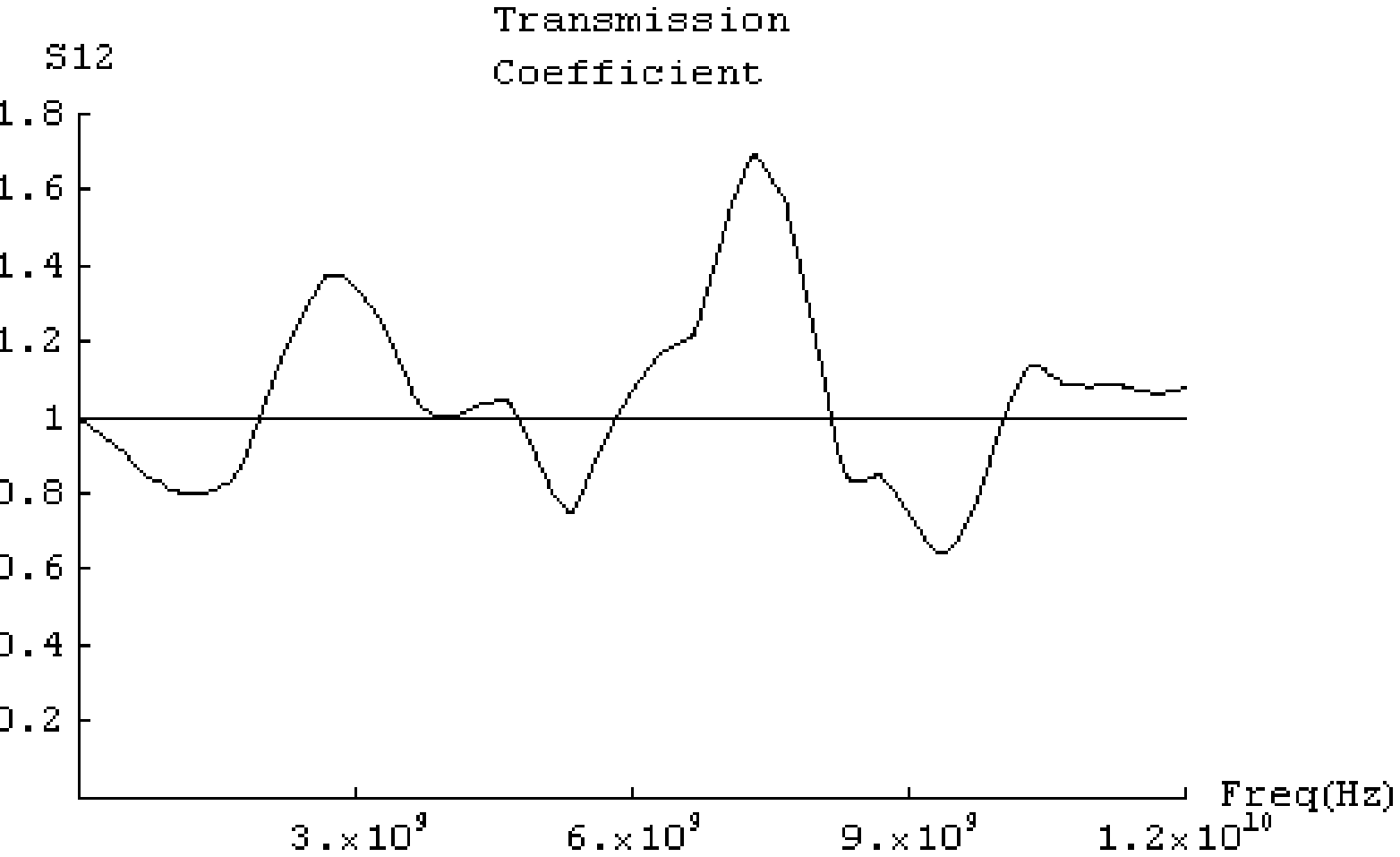}
\caption{\label{fig:S12}$|S_{12}|$}
\end{figure}
\begin{figure}
\includegraphics*[scale=0.4]{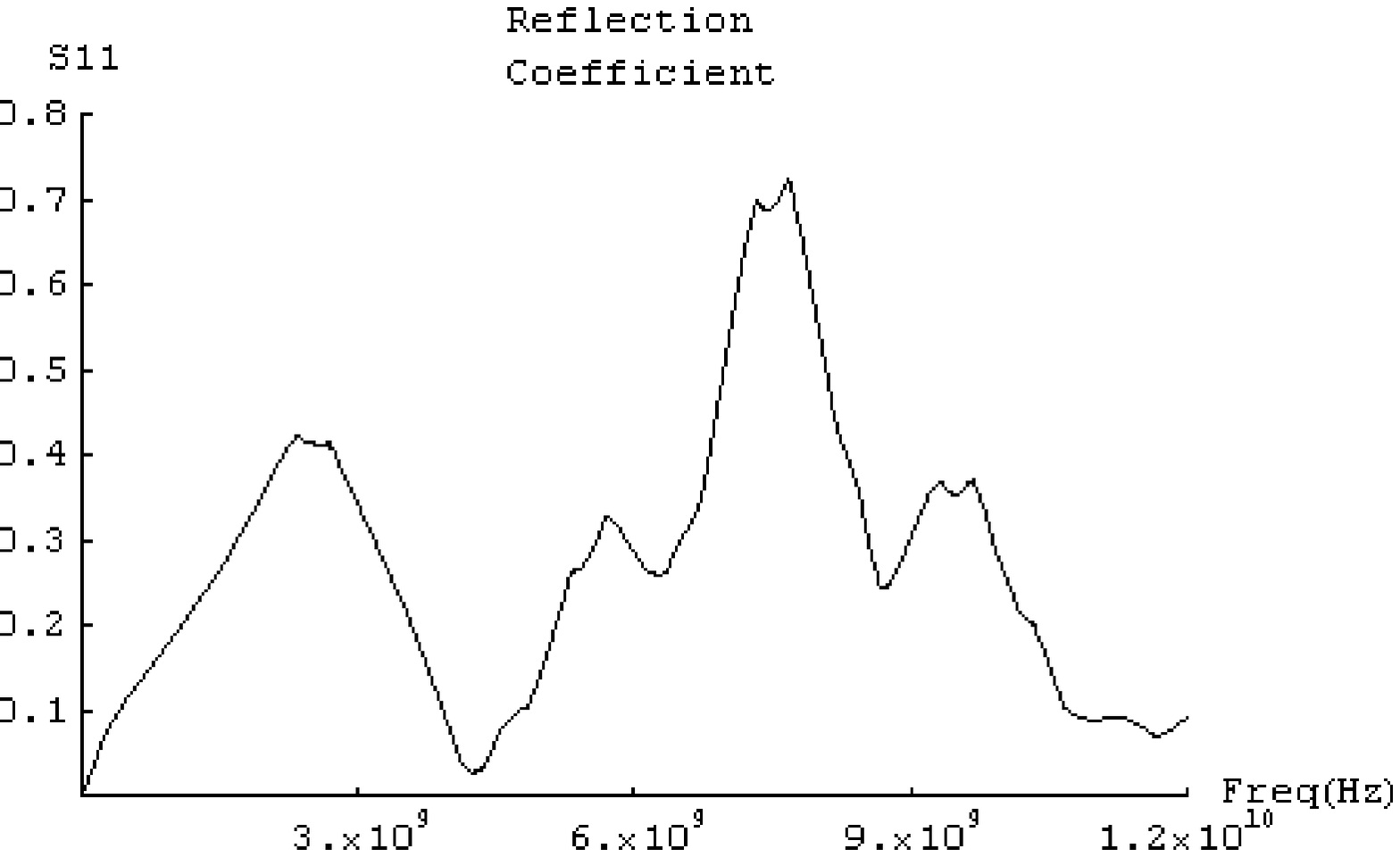}
\caption{\label{fig:S11}$|S_{11}|$}
\end{figure}
Using the above complex-valued reflection function we can validate this approach by simulating the reflection of an ideal step function. This should match the original $\rho$ data. Figure \ref{fig:simTDR} shows an almost exact fit with the experimental data (dashed and offset by 0.05 from the simulation).
\begin{figure}
\includegraphics*[scale=0.4]{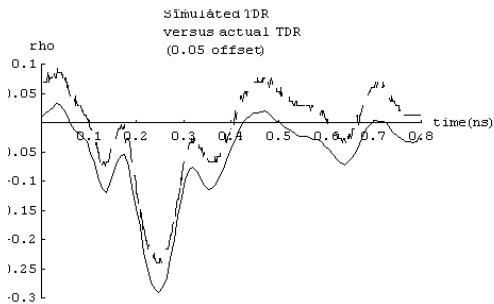}
\caption{\label{fig:simTDR}Simulated TDR vs real TDR}
\end{figure}

\section{Cables}
The next element to consider for copper interconnect is the cable. We need to determine an accurate model that captures the experimental data.

\subsection{Experimental data}
Figure \ref{fig:cableTDR} shows the experimental data for a 4.5 meter D7 cable with a shorted end. This data was taken by using TDR. The time axis measures roundtrip delay times. The ringing waveform on the left side is due to the fixture used to couple the TDR equipment to the cable and is excised from successive analysis. There are several things to note about this figure. First there is a linear rise in resistance as time progresses. This is due to the DC resistance of the cable. The TDR pulse scatters off this incremental resistance and produces a gradual series of return signals that add up to the observed ramp. The next feature is the abrupt drop at about 50 ns time delay. This is where the TDR pulse hits the short at the end of the cable. For a very short cable this drop would be almost perfectly vertical. For long cables the return pulse has been modified by the cable imperfections both on the outward trip and on the return trip. By analyzing the amplitude and phase changes long cables apply to each Fourier component we can derive a very accurate model of the cable's distributed parameters, e.g., loss per unit length, incremental inductance, etc.
\begin{figure}
\includegraphics*[scale=0.4]{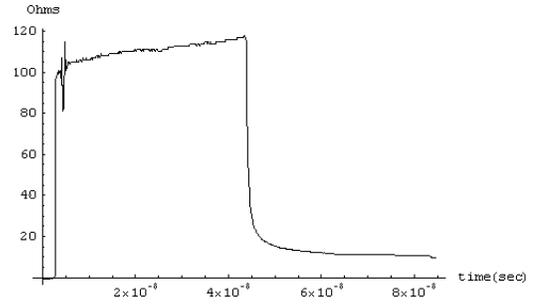}
\caption{\label{fig:cableTDR}TDR data for 4.5 meter cable}
\end{figure}

\subsection{Cable equations}
Two-conductor cables operating purely in differential mode are described by the partial differential equations
\begin{equation}
\partial_z v(z,t) = -r \, i(z,t) - l  \, \partial_t i(z,t)
\end{equation}
\begin{equation}
\partial_z i(z,t) = -g \, v(z,t) - c \, \partial_t v(z,t)
\end{equation}
where $r$ is the resistance of the cable per unit length, $c$ is the capacitance, and $l$ is the inductance. More generally we can make $r$ frequency dependent. $g$ is the transverse conductance between the two conductors, the $g$ term is neglected by assuming there are no dielectric losses. As a subtle point, $c$ and $l$ are assumed to be determined by the cable's wire diameters, i.e., we are assuming perfect conductors and perfect shielding in the above partial differential equations. We will correct for the effects of skin depth on these equations later.

We need to solve these equations for one-way energy flow on an infinitely long cable. First we assume a harmonic time dependence
\begin{equation}
v'(z)e^{-\imath \,t\,\omega} = -\left( r\,i(z)e^{-\imath \,t\,\omega } \right)+ \imath \,l\,\omega \,i(z)e^{-\imath \,t\,\omega }
\end{equation}
\begin{equation}
i'(z)e^{-\imath \,t\,\omega } = \imath \,c\,\omega \,v(z)e^{-\imath \,t\,\omega }
\end{equation}
Removing the common time dependence
\begin{eqnarray}
v'(z) & = & -\left( r\,i(z) \right)  + \imath \,l\,\omega \,i(z),\; \nonumber \\
i'(z) & = & \imath \,c\,\omega \,v(z)
\end{eqnarray}

Now we solve these differential equations
\begin{widetext}
\begin{equation}
{{v(z)} = {\frac{\left( 1 + e^{2\,z\,{\sqrt{-\imag \,c\,r\,\omega  - c\,l\,{\omega }^2}}} \right) \,\Mfunction{C}(1)}
      {2\,e^{z\,{\sqrt{-\imag \,c\,r\,\omega  - c\,l\,{\omega }^2}}}} - 
     \frac{\left( -1 + e^{2\,z\,{\sqrt{-\imag \,c\,r\,\omega  - c\,l\,{\omega }^2}}} \right) \,
        \left( r - \imag \,l\,\omega  \right) \,\Mfunction{C}(2)}{2\,e^{z\,{\sqrt{-\imag \,c\,r\,\omega  - c\,l\,{\omega }^2}}}\,
        {\sqrt{-\imag \,c\,r\,\omega  - c\,l\,{\omega }^2}}}}}
\end{equation}
\begin{equation}
{{i(z)} = {\frac{\frac{\imag }{2}\,c\,\left( -1 + e^{2\,z\,{\sqrt{-\imag \,c\,r\,\omega  - c\,l\,{\omega }^2}}} \right) \,
        \omega \,\Mfunction{C}(1)}{e^{z\,{\sqrt{-\imag \,c\,r\,\omega  - c\,l\,{\omega }^2}}}\,
        {\sqrt{-\imag \,c\,r\,\omega  - c\,l\,{\omega }^2}}} + 
     \frac{\left( 1 + e^{2\,z\,{\sqrt{-\imag \,c\,r\,\omega  - c\,l\,{\omega }^2}}} \right) \,\Mfunction{C}(2)}
      {2\,e^{z\,{\sqrt{-\imag \,c\,r\,\omega  - c\,l\,{\omega }^2}}}}}}
\end{equation}
Collecting coefficients between outgoing and ingoing waves
\begin{equation}
i(z) = \frac{\frac{\frac{-\imag }{2}\,c\,\omega \,\Mfunction{C}(1)}{{\sqrt{-\imag \,c\,r\,\omega  - c\,l\,{\omega }^2}}} + 
     \frac{\Mfunction{C}(2)}{2}}{e^{z\,{\sqrt{-\imag \,c\,r\,\omega  - c\,l\,{\omega }^2}}}} + 
  e^{z\,{\sqrt{-\imag \,c\,r\,\omega  - c\,l\,{\omega }^2}}}\,
   \left( \frac{\frac{\imag }{2}\,c\,\omega \,\Mfunction{C}(1)}{{\sqrt{-\imag \,c\,r\,\omega  - c\,l\,{\omega }^2}}} + 
     \frac{\Mfunction{C}(2)}{2} \right)
\end{equation}
\begin{eqnarray}
v(z)& = &\frac{\frac{\Mfunction{C}(1)}{2} + \frac{r\,\Mfunction{C}(2)}{2\,{\sqrt{-\imag \,c\,r\,\omega  - c\,l\,{\omega }^2}}} - 
     \frac{\frac{\imag }{2}\,l\,\omega \,\Mfunction{C}(2)}{{\sqrt{-\imag \,c\,r\,\omega  - c\,l\,{\omega }^2}}}}{e^
     {z\,{\sqrt{-\imag \,c\,r\,\omega  - c\,l\,{\omega }^2}}}} \\
& + &  e^{z\,{\sqrt{-\imag \,c\,r\,\omega  - c\,l\,{\omega }^2}}}\,
   \left( \frac{\Mfunction{C}(1)}{2} - \frac{r\,\Mfunction{C}(2)}{2\,{\sqrt{-\imag \,c\,r\,\omega  - c\,l\,{\omega }^2}}} + 
     \frac{\frac{\imag }{2}\,l\,\omega \,\Mfunction{C}(2)}{{\sqrt{-\imag \,c\,r\,\omega  - c\,l\,{\omega }^2}}} 
    \right) \nonumber 
\end{eqnarray}
\end{widetext}

We now define C(2) to eliminate the incoming wave
\begin{equation}
 {{\Mfunction{C}(2)} =
     {\frac{\imag \,c\,\omega \,\Mfunction{C}(1)}
       {{\sqrt{-\imag \,\Mfunction{c}\,r\,\omega  - 
            \Mfunction{c}\,l\,{\omega }^2}}}}}
\end{equation} 
We have
\begin{equation}
v(z) = e^{z\,{\sqrt{-\left( c\,\omega \,\left( \imag \,r + l\,\omega  \right) 
            \right) }}}\,\Mfunction{C}(1)
\end{equation} 
\begin{equation}
i(z) =  \frac{\imag \,c\,e^{z\,{\sqrt{-\imag \,c\,r\,\omega  - c\,l\,{\omega }^2}}}\,
     \omega \,\Mfunction{C}(1)}{{\sqrt{-\imag \,c\,r\,\omega  - 
        c\,l\,{\omega }^2}}}
\end{equation}

We now derive the cable's attenuation factor. Focusing upon $v(z)$ and setting $C(1) = 1$ we have
\begin{equation}
v(z) =  e^{z\,{\sqrt{-\left( c\,\omega \,\left( \imag \,r + l\,\omega  \right) 
           \right) }}}
\end{equation}
To further simplify this equation we note that $r$ is a small impedance compared to the inductive impedance of the cable, for all high enough frequencies. By expanding the square root around the large inductive term and dropping the imaginary phase-determining part, we derive the loss per unit length of the cable.
\begin{equation}
\mbox{loss} = e^{\frac{-\left( {\sqrt{c}}\,r\,z \right) }{2\,{\sqrt{l}}}}
\end{equation}
In log terms
\begin{equation}
\mbox{logLoss} = \frac{-\left( {\sqrt{c}}\,r\,z \right) }{2\,{\sqrt{l}}}
\end{equation}
Which results in
\begin{equation}
r = \frac{-2\,{\sqrt{l}}\,\Mvariable{\mbox{logLoss}}}{{\sqrt{c}}\,z}
\end{equation}

This formula allows us to determine $r$ as a function of frequency from the experimentally measured cable loss versus frequency relationship. We set $z$ to the measured length of the cable, times 2 to account for the roundtrip, to derive the $r$ per unit length values. This frequency-dependent $r$ function produces a much more accurate model for real cables then a simple skin depth effect model.

In Appendix \ref{App:skin} we show that the skin effect produces both a resistive term, due to losses in the wire from the finite conductivity of the metal, and an additional inductive term, due to the finite penetration of magnetic flux into the wire. After we measure the real part of r from the experimental cable data we add back the skin effect's additional inductive effects by using the "complex" resistance value: $r_{\mbox{extracted}}(1 - \imag)$.

\subsection{Experimental cable results}
Figure \ref{fig:cableTDR} shows the full experimental data set. The most interesting part of the data set is the sharp return step. To isolate just this element we window the data set with a gaussian envelope, centered upon the step, but broad enough to preserve the lower frequency components. Figure \ref{fig:WinCableTDR} shows the result.
\begin{figure}
\includegraphics*[scale=0.4]{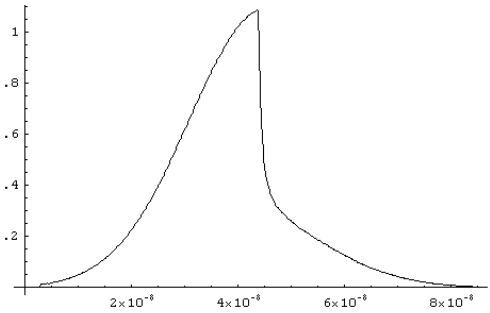}
\caption{\label{fig:WinCableTDR}Windowed cable data}
\end{figure}

To compare the experimental data with a known waveform we do the same gaussian windowing with an ideal return pulse that has been scaled and positioned to best match the experimental data. Figure \ref{fig:WinIdealTDR} shows this waveform.
\begin{figure}
\includegraphics*[scale=0.4]{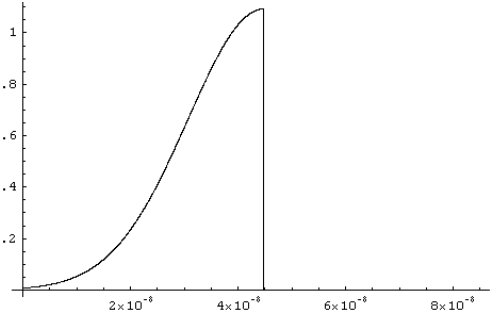}
\caption{\label{fig:WinIdealTDR}Windowed ideal model}
\end{figure}

After Fourier transforming both of these windowed pulses we divide the experimental Fourier components by the ideal Fourier components, frequency by frequency. Figure \ref{fig:cableLoss} shows the loss function of the measured cable.
\begin{figure}
\includegraphics*[scale=0.4]{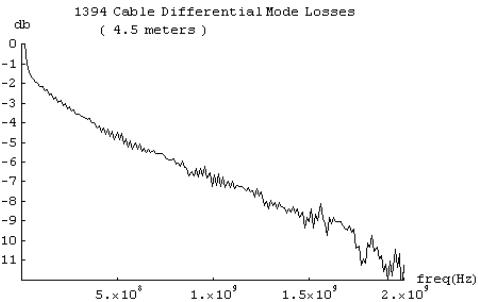}
\caption{\label{fig:cableLoss}Cable losses}
\end{figure}

Using the formula above we derive the cable's measured resistance factor shown in Figure \ref{fig:cableR}.
\begin{figure}
\includegraphics*[scale=0.4]{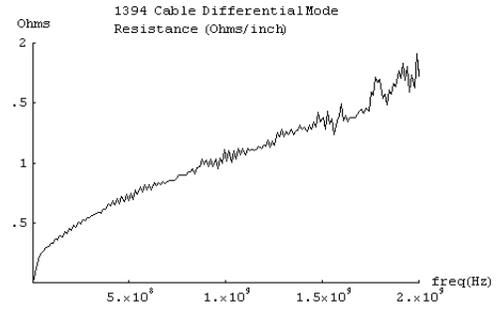}
\caption{\label{fig:cableR}Cable resistance}
\end{figure}

Figure \ref{fig:cableCmp} shows a close-up of the experimental TDR's step edge versus a simulation based upon the mathematical methods we've developed above. The bold plot is the data. We see reasonable agreement.
\begin{figure}
\includegraphics*[scale=0.4]{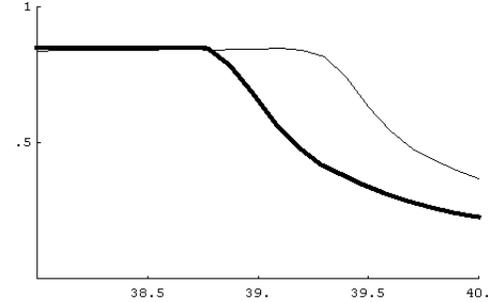}
\caption{\label{fig:cableCmp}Experimental edge (bold) compared with the modeled edge}
\end{figure}

\section{Connector plus cable systems}
Given two connectors and a length of connecting cable we have an infinity of reflections which sum up to a net transfer function defined by
\begin{equation}
S_\infty = \frac{H(\omega) T^2(\omega)}{1-R^2(\omega)H^2(\omega)}
\end{equation}
$H(\omega)$ is the cable transfer function, $T(\omega)$ is the transmission coefficient for the connectors, and $R(\omega)$ is the reflection coefficient for the connectors.
\begin{figure}
\includegraphics*[scale=0.4]{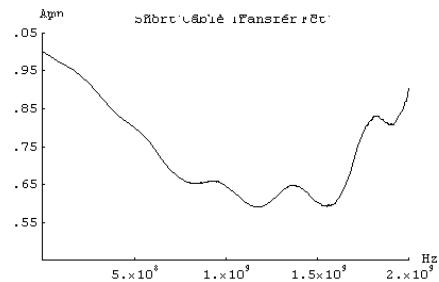}
\caption{\label{fig:shortConn}Two connectors and a short length of cable}
\end{figure}

Using the connector and cable models we've derived we have the composite transfer funcition shown in Figure \ref{fig:shortConn}, where we've made the cable very short. Signficant resonances are visible.

\section{Eye diagrams}

An important quantity to determine for a copper interconnect system is the eye diagram. We synthesize a stressful set of data transitions, apply them to the copper system transfer function, and see how the output pattern looks. The basic technique is to convert a bit pattern to a pulse pattern with the right timing, e.g., rise/fall times, etc. We produce a discrete, time-sampled waveform from the pulse pattern, Fourier transform it, multiply by the transfer function, then inverse Fourier transform to get the output waveform in time.

We use the following bit pattern to test the cable/connectors system: 11000001 01001111 10101000 00000000 00000000 00000000 01011111 11111111 11111111 11111110 $\cdots$ repeating.

\begin{figure}
\includegraphics*[scale=0.4]{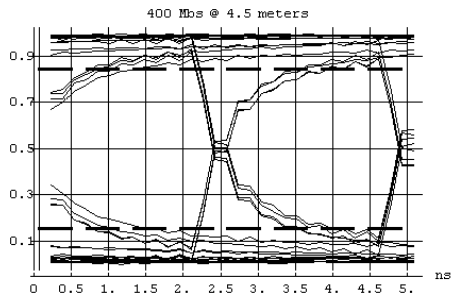}
\caption{\label{fig:eye}400 Mbps eye diagram}
\end{figure}
Figure \ref{fig:eye} shows the result of this analysis. We have sliced the output waveform into 1 bit-time length segments and then aligned and superimposed these slices. The result of the simulations is a very accurate model of experimentally determined eye diagrams (not shown here). Eye diagram simulations permit the determination of useful things like total system margins, jitter budgets, etc.

\section{Conclusions}
TDR Oscilloscopes are readily available pieces of equipment. By using TDR techniques along with the mathematical methods described in this paper it is possible to characterize and model many of the interesting features of a copper interconnect system.

\appendix
\section{\label{App:analytic}The Analytic signal}   
We have used analytic signals (see Gabor\cite{gabor}) to determine the phases of complex-valued transfer functions when all we originally had was the real part of the function. In this appendix the theory behind analytic signals is discussed.

The essential assumption behind analytic signals is that we are dealing with a linear system that obeys causality, i.e., responses occur after inputs in time. The most general way to express this is by the integral equation:
\begin{equation}
\phi(t) = \int d\acute{t} K(t-\acute{t})g(\acute{t})
\end{equation}
$K(t-\acute{t})$ is the Green's function for the system (e.g., cable, connector), $g(\acute{t})$ in the input (e.g., incident voltage waveform), and $\phi(t)$ is the system's response (e.g., reflected, transmitted waves). This integral equation comprehends all the time lags of the linear system. We require causality which means that $K(t) = 0$ for $t < 0$. The function of interest for our system is the Fourier transform of $K(t)$ and the condition $K(t) = 0$ for $t < 0$ implies that the Fourier transform
\begin{equation}
f(z) = \int dt e^{i t z} K(t)
\end{equation}
of $K(t)$ has no singularities for $\mbox{Im}(z) > 0$, and that $f(z)\to 0$ for $\vert z \vert \to \infty$ in the upper half z-plane.

We now apply Cauchy's formula to a contour consisting of the real axis and a large upper semicircle, see Figure \ref{fig:contour}.
\begin{figure}
\includegraphics*[scale=0.4]{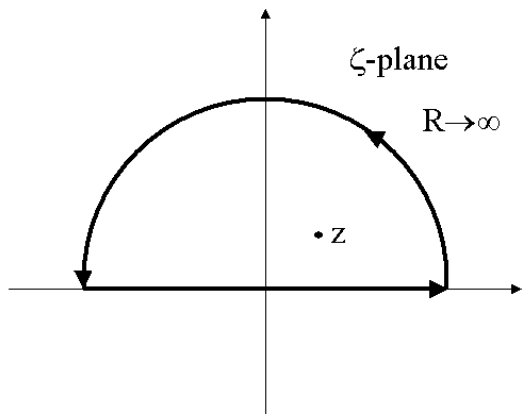}
\caption{\label{fig:contour}Integration contour}\end{figure}
By the above condition on $f(z)$ we have the contribution from the semicircle to the line integral along the contour going to zero. This implies that Cauchy's formula reduces to
\begin{equation}
f(z) = \frac{1}{2 i \pi} \int_{\mbox{real axis}}\frac{f(\zeta)}{\zeta - z}d\zeta, \;\mbox{Im}z > 0
\end{equation}
The function of physical interest is $f(z)$ on the real axis (i.e., real-valued frequencies). Define the function $F(x)$ to be
\begin{equation}
F(x) \equiv \lim_{\epsilon \to 0} f(x + i \epsilon)
\end{equation}
where $\epsilon$ is positive. Then it follows that
\begin{eqnarray}
2 i \pi F(x) & = & \lim _{\epsilon \to 0} \int _{- \infty} ^{\infty} \frac{F(\acute{x})}{\acute{x} - x - i \epsilon} d \acute{x} \nonumber \\
& = & P \int _{- \infty} ^{\infty} \frac{F(\acute{x})}{\acute{x} - x} d \acute{x} + i \pi F(x) \nonumber \\
&&
\end{eqnarray}
where $P$ designates the Cauchy principal value. Thus
\begin{equation}
F(x) = \frac{1}{i \pi} P \int _{- \infty} ^{\infty} \frac{F(\acute{x})}{\acute{x} - x} d \acute{x}
\end{equation}
\begin{equation}
\mbox{Re} F(x) = \frac{1}{\pi} P \int _{- \infty} ^{\infty} \frac{\mbox{Im}F(\acute{x})}{\acute{x} - x} d \acute{x},
\end{equation}
\begin{equation}
\mbox{Im} F(x) = \frac{-1}{\pi} P \int _{- \infty} ^{\infty} \frac{\mbox{Re}F(\acute{x})}{\acute{x} - x} d \acute{x},
\end{equation}

These are dispersion relations for the function of physical interest $F(x)$ and the last equation clearly shows how to calculate the imaginary component of any transfer function from knowledge of its real part. This equation is the Hilbert transform. Note in our application x is frequency.

As a simple example let us consider a simple low-pass RC circuit with unit R and unit C in series. $V_{\mbox{in}}$ is applied across the series combination and $V_{\mbox{out}}$ is measured across the capacitor. The complex transfer function is
\begin{equation}
\frac{V_{out}}{V_{in}} = \frac{\frac{-1}{i \omega}}{1-\frac{1}{i \omega}} = \frac{1}{1 - i \omega}
\end{equation}
which can be expressed in real and imaginary parts explicitly as
\begin{equation}
\frac{V_{out}}{V_{in}}  = \frac{1}{1 + \omega^2} + \frac{i \omega}{1 + \omega^2}
\end{equation}
\begin{equation}
\mbox{real part} = \frac{1}{1+\omega^2}
\end{equation}

We now assume that all we know is the real part of the transfer function. The goal is to use analytic signal theory to derive the correct imaginary part.
Negative frequencies don't make sense physically in this situation but this form of $F$ possesses the symmetry:
\begin{equation}
f(-z) = +f^*(z^*)
\end{equation}
where * denotes complex conjugation. This implies the relationship
\begin{eqnarray}
F(x) & = & \frac{1}{i \pi} P \int _{\infty} ^0 \frac{F^*(\acute{x})}{\acute{x} + x} d \acute{x} \nonumber \\
&& + \frac{1}{i \pi} P \int ^{\infty} _0 \frac{F(\acute{x})}{\acute{x} - x} d \acute{x}
\end{eqnarray}
For the imaginary part
\begin{eqnarray}
\mbox{Im}F(x) & = & \frac{-1}{\pi} P \int ^{\infty} _0 \frac{-\mbox{Re}F(\acute{x})}{\acute{x} + x} d \acute{x} \nonumber \\
&& + \frac{1}{\pi} P \int ^{\infty} _0 \frac{-\mbox{Re}F(\acute{x})}{\acute{x} - x} d \acute{x} \nonumber \\
&&
\end{eqnarray}
which simplifies to
\begin{equation}
\mbox{Im}F(x) = \frac{-2}{\pi} P \int ^{\infty} _0 \frac{x \mbox{Re}F(\acute{x})}{\acute{x}^2 - x^2} d \acute{x}
\end{equation}

Putting in the explicit value of F
\begin{eqnarray}
\mbox{Im}F(\omega) & = & \frac{-2}{\pi} P \int ^{\infty} _0 \frac{\frac{\omega}{1+\acute{\omega^2}}}{\acute{\omega}^2 - \omega^2} d \acute{\omega} \nonumber \\
&= & \frac{-2}{\pi} P \int ^{\infty} _0 \frac{\omega}{(\acute{\omega}^2 - \omega^2)(1+\acute{\omega}^2)} d \acute{\omega} \nonumber \\
&&
\end{eqnarray}
Given the definition of the Principal part:
\begin{widetext}
\begin{equation}
P \int ^{\infty} _0 \frac{\omega}{(\acute{\omega}^2 - \omega^2)(1+\acute{\omega}^2)} d \acute{\omega} = 
 \lim _{\delta \to 0}\Bigl[ \int ^{\omega - \delta} _0 \frac{\omega}{(\acute{\omega}^2
 - \omega^2)(1+\acute{\omega}^2)} d \acute{\omega} + \int _{\omega + \delta} ^\infty \frac{\omega}{(\acute{\omega}^2 - \omega^2)(1+\acute{\omega}^2)} d \acute{\omega} \Bigr]
\end{equation}
\end{widetext}
where $\delta$ is positive. Performing the integrations
\begin{equation}
\mbox{Im} F(x) = \frac{\omega}{1+\omega^2}
\end{equation}
which is exactly the imaginary part of the transfer function.

\section{Skin effect} \label{App:skin}   
Accurate prediction of the behavior of copper transmission lines requires proper modeling of the losses and incremental inductance factors caused by the skin effect. Here we treat the skin effect without the usual assumption of spatially constant conductivity.

Following Jackson\cite{jackson}we restrict Maxwell's equations to weak fields, linear, isotropic-response media, and the quasi-static regime -- where we neglect the displacement current contribution to the magnetic field.
\begin{eqnarray}
\mbox{\boldmath $\nabla \times H$} = \mbox{\boldmath $J$} & & \nonumber \\
\mbox{\boldmath $\nabla \bullet B$} = 0 & & \nonumber \\
\mbox{\boldmath $\nabla \times E + \frac{\partial B}{\partial t} = 0$} & & \nonumber \\
\mbox{\boldmath $J = \sigma E$}& & \nonumber \\
\mbox{\boldmath $B$} = \mu \mbox{\boldmath $H$}& &
\end{eqnarray}

With the definition of the vector potential, $\mbox{\boldmath $B = \nabla \times A$}$, Faraday's law shows that the curl of $\mbox{\boldmath $E + \partial A / \partial t$}$ vanishes. This implies that we can write $\mbox{\boldmath $E = -\partial A / \partial t - \nabla$} \Phi$. Assuming negligible free charge and that the time-varying $\mbox{\boldmath $B$}$ is the sole source of the electric field, we may set the scalar potential $\Phi$ to zero and have $\mbox{\boldmath $E = -\partial A / \partial t$}$. Note that we have the subsidiary conditions $\mbox{\boldmath $\nabla \bullet E = 0$}$ and $\mbox{\boldmath $\nabla \bullet A = 0$}$. For media of uniform, frequency-independent permeability $\mu$, Ampere's law can be written $\mbox{\boldmath $\nabla \times B = $} \mu \mbox{\boldmath $J$} = \mu \sigma \mbox{\boldmath $E$}$. Elimination of $\mbox{\boldmath $B$}$ and $\mbox{\boldmath $E$}$ in favor of $\mbox{\boldmath $A$}$ and use of the vector identity, $\mbox{\boldmath $\nabla \times \nabla \times A = \nabla(\nabla \bullet A)$} - \nabla^2 \mbox{\boldmath $A$}$, yields the diffusion equation for the vector potential
\begin{equation}
\mbox{\boldmath $\nabla^2 A$} = \mu \sigma \frac{\partial \mbox{\boldmath $A$}}{\partial t}
\end{equation}

This equation, which obviously also holds for the electric field is valid for spatially varying, but frequency-independent conductivity. If the conductivity is constant in space it follows that the magnetic induction and the current density also satisfy the same diffusion equation. This situation gives the usual skin depth effect. Constant conductivity is a poor approximation for real wires which are subject to varying alloy concentrations and stresses due to processing. The goal of this appendix is to derive the overall relationship between the driving current and electric fields in realistic wires with spatially varying conductivity. 

To proceed we assume cylindrical coordinates along the wire, harmonic time dependence, and use complex phasor notation, $\mbox{\boldmath $A$}(\rho, \phi, z, t) = \mbox{\boldmath $A$}(\rho, \phi, z)e^{- \imath \omega t}$. After dropping the common exponential factor
\begin{equation}
\mbox{\boldmath $\nabla^2 A$}(\rho, \phi, z) = -\imath \omega \mu \sigma(\rho, \phi, z)\mbox{\boldmath $A$}(\rho, \phi, z)
\end{equation}

Assuming the wire diameter is much smaller that the wavelengths of interest, we drop the $\phi$ dependence. Given that a constant amount of current is flowing along the $z$ axis of the wire we see that the curl of $\mbox{\boldmath $A$}$ at the surface of the wire must be constant, independent of $z$. This follows from the relationship between $\mbox{\boldmath $B$}$ and $\mbox{\boldmath $A$}$ and Gauss's law relating the magnetic field around a contour and the enclosed current. $\mbox{\boldmath $E$}$ is very small in the deep interior of the wire at high enough frequencies, thus $\mbox{\boldmath $A$}$ is very small at the center of the wire. We also assume that $\mbox{\boldmath $E$}$ is parallel to the wire at its surface, given there is no free charge on the surface. The only non-trivial dependence is that of $\sigma$ and $\mbox{\boldmath $A$}$ on $\rho$ and $z$. We ignore $\sigma$'s and $\mbox{\boldmath $A$}$'s dependence upon $z$ since each different position along the wire will have an independent voltage drop due to the very small skin depth expected at interesting frequencies -- the net voltage to driving current ratio for the entire cable is simply the average along the wire's length. Thus we assume a uniform and constant dependence along $z$ for the diffusion equation. We now have a set of equations to solve
\begin{widetext}
\begin{equation}
\nabla^2 A_z(\rho,z)= \frac{1}{\rho} \frac{\partial}{\partial \rho} (\rho \frac{\partial A_z}{\partial \rho}) + (\frac{\partial^2 A_z}{\partial z^2} \rightarrow 0) =  -\imath \omega \mu \sigma(\rho) A_z(\rho),
\end{equation}
\begin{equation}
\nabla^2 A_\rho(\rho,z)= \frac{1}{\rho} \frac{\partial}{\partial \rho} (\rho \frac{\partial A_\rho}{\partial \rho}) + (\frac{\partial^2 A_\rho}{\partial z^2} \rightarrow 0) =  -\imath \omega \mu \sigma(\rho) A_\rho(\rho),
\end{equation}
\begin{equation}
\mbox{\boldmath $\nabla \times A \cong e_\phi$}((\frac{\partial A_\rho}{\partial z} \rightarrow 0) - \frac{\partial A_z(\rho)}{\partial \rho}) \mid_{\rho=r_{\mbox{wire}}} =  \mbox{\boldmath $B$}_{\mbox{surface}},
\end{equation}
\end{widetext}
\begin{equation}
A_\rho(0) = A_z(0) \cong 0,
\end{equation}
\begin{equation}
A_\rho(r_{\mbox{wire}}) = 0
\end{equation}
$A_\rho(r_{\mbox{wire}}) = 0$ -- since the electric field at the surface of the wire is strictly oriented along the wire's axis. Given the boundary conditions we see the $A_\rho$ always remains zero. There are no terms in the partial differential equation for $A_\rho$ that can drive a non-zero value anywhere, $A_\rho$ being decoupled from $A_z$. All that remains is
\begin{equation}
\nabla^2 A_z(\rho,z)= \frac{1}{\rho} \frac{\partial}{\partial \rho} (\rho \frac{\partial A_z}{\partial \rho}) =  -\imath \omega \mu \sigma(\rho) A_z(\rho),
\end{equation}
\begin{equation}
\mbox{\boldmath $\nabla \times A \cong -e_\phi$}\frac{\partial A_z(\rho)}{\partial \rho} \mid_{\rho=r_{\mbox{wire}}} =  \mbox{\boldmath $B$}_{\mbox{surface}},
\end{equation}
\begin{equation}
A_z(0) \cong 0
\end{equation}

As a first step we solve these equations for spatially constant $\sigma$
\begin{equation}
A_z(\rho) = J_0(-\sqrt[4]{-1} \sqrt{\mu \sigma \omega} \rho) c_1 + Y_0(-\sqrt[4]{-1} \sqrt{\mu\sigma \omega} \rho) c_2
\end{equation}
$Y_0(0) = - \infty$, thus to preserve the small value of $A_z$ at the origen of the wire we set $c_2$ to zero.
\begin{equation}
A_z(\rho) = J_0(-\sqrt[4]{-1} \sqrt{\mu \sigma \omega} \rho) c_1,
\end{equation}
\begin{equation}
\frac{\partial A_z(\rho)}{\partial \rho} = J_1(-\sqrt[4]{-1} \sqrt{\mu \sigma \omega}\rho) c_1 \sqrt[4]{-1} \sqrt{\mu \sigma \omega}
\end{equation}
Setting the derivative at $\rho = r_{\mbox{wire}}$ equal to the surface magnetic field we have the full solution
\begin{widetext}
\begin{equation}
A_z(\rho) = -\frac{B_{\mbox{surface}}}{\sqrt[4]{-1} \sqrt{\mu \sigma \omega}
J_1(-\sqrt[4]{-1} \sqrt{\mu  \sigma \omega} r_{\mbox{wire}})}J_0(-\sqrt[4]{-1} \sqrt{\mu \sigma \omega} \rho),
\end{equation}
\begin{equation}
\frac{\partial A_z(\rho)}{\partial \rho} = -\frac{B_{\mbox{surface}}}{J_1(-\sqrt[4]{-1} \sqrt{\mu  \sigma \omega} r_{\mbox{wire}})}J_1(-\sqrt[4]{-1} \sqrt{\mu \sigma \omega} \rho)
\end{equation}
\end{widetext}
which describes a thin layer of excitation diminishing rapidly in the interior of the wire.

We now want to determine the relationship between the electric field and the magnetic induction at the wire's surface. The magnetic induction at the surface is a direct function of the driving current inside the wire.
\begin{widetext}
\begin{equation}
E_z(\rho) = - \partial A_z(\rho) / \partial t = \imath \omega A_z(
\rho) = -\frac{\imath \omega B_{\mbox{surface}}}{\sqrt[4]{-1} \sqrt{\mu \sigma \omega}
J_1(-\sqrt[4]{-1} \sqrt{\mu  \sigma \omega} r_{\mbox{wire}})}J_0(-\sqrt[4]{-1} \sqrt{\mu \sigma \omega} \rho),
\end{equation}
\end{widetext}
\begin{equation}
\mbox{\boldmath $B = $}\mu \mbox{\boldmath $H$},
\end{equation}
\begin{widetext}
\begin{equation}
\mbox{\boldmath $B = \nabla \times A = e_\phi$}((\frac{\partial A_\rho(\rho)}{\partial z} \rightarrow 0) - \frac{\partial A_z(\rho)}{\partial \rho})
\end{equation}
\end{widetext}
we get
\begin{widetext}
\begin{equation}
H_\phi = \frac{B_\phi}{\mu} = -\frac{1}{\mu} \frac{\partial A_z(\rho)}{\partial \rho} = \frac{B_{\mbox{surface}}}{\mu J_1(-\sqrt[4]{-1} \sqrt{\mu  \sigma \omega} r_{\mbox{wire}})}J_1(-\sqrt[4]{-1} \sqrt{\mu \sigma \omega} \rho)
\end{equation}
\end{widetext}
which is proportional to the total driving current in the wire. Consequently
\begin{equation}
\frac{E_z(\rho)}{H_\phi(\rho)} = -\frac{\sqrt[4]{-1} \sqrt{\mu \omega}J_0(-\sqrt[4]{-1} \sqrt{\mu \sigma \omega} \rho)}{\sqrt{\sigma}J_1(-\sqrt[4]{-1} \sqrt{\mu \sigma \omega} \rho)}
\end{equation}
The two Bessel functions grow at the same rate for large arguments and their ratio $\rightarrow \imath$. Consequently we have the wire surface result
\begin{eqnarray}
\frac{E_z(r_{\mbox{wire}})}{H_\phi(r_{\mbox{wire}})} & = & -\frac{\sqrt[4]{-1} \sqrt{\mu \omega}\imath}{\sqrt{\sigma}} \nonumber \\
& = & \frac{(1-\imath)}{\sqrt{2}} \times \mbox{real parameters} \nonumber \\
&&
\end{eqnarray}
Thus the electric field along the wire has a constant phase shift relative to the driving current. This formula justifies the use of the complex constant times the measured high frequency resistance in the transmission line equations -- again for constant $\sigma$.

For the case of spatially varying $\sigma$ we simplify matters by assuming a semi-infinite slab of material with the vacuum-matter interface at $x = 0$, extending along $y$ and $z$ to infinity. Assume the applied magnetic induction is parallel to the $y$ axis. This implies the vector potential only has a $z$ component. Because skin depths are much smaller than wire radii for most problems of interest -- a 1D approach is reasonable.
\begin{equation}
\nabla^2 A_z(x)= \frac{\partial^2 A_z}{\partial x^2} = -\imath \omega \mu \sigma(x) A_z(x),
\end{equation}
\begin{equation}
\mbox{\boldmath $\nabla \times A = e_y$} \frac{\partial A_z(x)}{\partial x}\mid_{x=0} = \mbox{\boldmath $B_{\mbox{surface}}$}
\end{equation}
\begin{equation}
A_z(x \rightarrow \infty) = 0
\end{equation}

The simplest solution is to assume a Taylor's series expansion for all the terms in the equation. To simplify further we assume that the spatially varying conductivity is a small perturbation added to a constant conductivity.
\begin{equation}
\sigma(x) = \sigma_0 + \epsilon \sigma_1(x), \epsilon \ll 1,
\end{equation}
\begin{equation}
A_z(x) = A_0(x) + \epsilon A_1(x) + \epsilon^2 A_2(x) + \cdots
\end{equation}
\begin{equation}
\frac{\partial^2 A_z(x)}{\partial x^2} =  \frac{\partial^2 A_0(x)}{\partial x^2} +  \epsilon \frac{\partial^2 A_1(x)}{\partial x^2} + \cdots
\end{equation}

Applying these series to the diffusion equation we separate out the various contributions to the different powers of $\epsilon$
\begin{equation}
\frac{\partial^2 A_0(x)}{\partial x^2} = -\imath \omega \mu \sigma_0 A_0(x)
\end{equation}
\begin{equation}
\frac{\partial^2 A_1(x)}{\partial x^2} = -\imath \omega \mu (\sigma_0 A_1(x) + \sigma_1(x) A_0(x))
\end{equation}
\begin{displaymath}
\cdots
\end{displaymath}

The solution to the $A_0$ equation with the correct boundary behavior at $x = \infty$ is
\begin{eqnarray}
A_0(x) & = & c_1 e^{-\sqrt{-\imath \omega \mu \sigma_0} x} \nonumber \\
& = & c_1 \sum_{n=0}^{\infty} \frac{(-\sqrt{-\imath \omega \mu \sigma_0} x)^n}{n!}
\end{eqnarray}
An important thing to notice is that the $A_0(x)$ solution multiplies $\sigma_1(x)$ in the equation determining $A_1(x)$. Since $A_0(x)$ is an exponentially decreasing function away from the surface, the variations in $\sigma_1(x)$ far away from the surface become irrelevant to $A_1(x)$. This explains why modeling the skin effect losses as a general function of frequency is necessary for accuracy. As the frequency changes, the skin depth changes -- thereby averaging the wire's conductivity over a different spatial scale.

To solve the $A_1$ equation we expand as follows
\begin{equation}
\sigma_1(x) = \sum_{n=0}^\infty s_n x^n
\end{equation}
\begin{equation}
A_1(x) = \sum_{n=0}^\infty a_n x^n
\end{equation}
\begin{eqnarray}
\Rightarrow \frac{\partial^2 A_1(x)}{\partial x^2} & = & \sum_{n=0}^\infty n(n-1)a_n x^{n-2} \nonumber \\
& =  & \sum_{n=0}^\infty (n+2)(n+1)a_{n+2} x^n \nonumber \\
&&
\end{eqnarray}
\begin{eqnarray}
\sigma_1(x) A_0(x)&& \nonumber \\
 & = & c_1 \sum_{n=0}^\infty \sum_{j=0}^n \frac{(-\sqrt{-\imath \omega \mu \sigma_0})^{n-j}}{(n-j)!} \times s_j x^n \nonumber \\
&&
\end{eqnarray}

Expanding the equations and separating by powers of $x$ we find an infinite series of equations for the progressively higher order coefficients. The first two equations are
\begin{eqnarray}
2 a_2 + \imath \omega \mu \sigma_0 a_0 + \imath \omega \mu c_1 \times s_0 && \nonumber \\
&& = 0 \nonumber \\
&&
\end{eqnarray}
\begin{eqnarray}
3 \times 2 a_3 + \imath \omega \mu \sigma_0 a_1 + \imath \omega \mu c_1 (-\sqrt{-\imath \omega \mu \sigma_0}) \times s_0 & & \nonumber \\
 + \imath \omega \mu c_1 \times s_1   =  0 && \nonumber \\
&&
\end{eqnarray}

Clearly $a_0$ and $a_1$ are unconstrained by the recursion relationship. In the case where all the $s_n = 0$ then $a_n = 0$ as well -- there being only a zero solution when there is no driving perturbation. Setting the first two coefficients to zero to match this limiting case, we see that the perturbed solution only has finite terms of quadratic and higher power. This implies that at the wire's surface ($x = 0$) there is no change in either the vector potential or its curl from the perturbation in $\sigma$. Thus the constant phase relationship between the electric field and the driving current in the wire remains as in the constant conductivity case.

\begin{acknowledgments}
The author would like to thank Steve Midford for providing the experimental data.
\end{acknowledgments}

\bibliography{copper}

\end{document}